\documentclass[a4paper]{amsart}
\usepackage{dsfont}
\usepackage{graphicx}
\usepackage[round]{natbib}

\newcommand{\cf}{\mathds{1}}  
\newcommand{\R}{\mathbb{R}}  
\newcommand{\Z}{\mathbb{Z}} 
\renewcommand{\P}{\mathbb{P}} 
\newcommand{\E}{\mathbb{E}} 
\newcommand{\bvc}[1]{\mathbf{#1}} 
\newcommand{\cF}{\mathcal{F}} 
\newcommand{\beq}[1]{\begin{equation}\label{#1}}
\newcommand{\eeq}{\end{equation}}
\newcommand{\beqn}[1]{\begin{equation} \nonumber}
\newcommand{\labeln}[1]{ \nonumber}
\newcommand{\dd}{\mathrm d}
\newcommand{\pdd}{\partial}

\begin{document}
% 
% \runningheads{J.~Br\"{o}cker}{Reliability of ensemble forecasting systems under serial dependence}
% 
\title{Assessing the reliability of ensemble forecasting systems under serial dependence}
\author{Jochen Br\"{o}cker}
\address{School of Mathematical and Physical Sciences, University of Reading, United Kingdom, \today}
\thanks{Fruitful discussions with Tobias Kuna are gratefully acknowledged.}
\begin{abstract}	
The problem of testing the reliability of ensemble forecasting systems is revisited. A
popular tool to assess the reliability of ensemble forecasting systems (for scalar
verifications) is the rank histogram; this histogram is expected to be more or less
flat, since for a reliable ensemble, the ranks are uniformly distributed among their
possible outcomes. Quantitative tests for flatness (e.g.\ Pearson's goodness--of--fit
test) have been suggested; without exception though, these tests assume the ranks to
be a sequence of independent random variables, which is not the case in general as can
be demonstrated with simple toy examples. In this paper, tests are developed that take
the temporal correlations between the ranks into account.  A refined analysis shows
that exploiting the reliability property, the ranks still exhibit strong decay of
correlations.  This property is key to the analysis, and the proposed tests are valid
for general ensemble forecasting systems with minimal extraneous assumptions. 

\end{abstract}
\maketitle  
% 
%%%%%%%%%%%%%%%%%%%%%%%%%%%% 
% 
\section{Introduction}
\label{sec:introduction}
A large proportion of environmental forecasting systems nowadays issue ensemble forecasts.
Such systems are used at major (national or international) weather centres, but may also form part of large scale research projects.
As with any forecasting system, there is a need to objectively assess the performance of ensemble forecasting systems.
Inasmuch as ensemble forecasts provide probabilistic information about the verification, such an assessment has to be statistical in character.
Several desirable (statistical) properties of ensemble (or more generally probabilistic) forecasting systems have been identified; see for instance~\citet{broecker09,broecker_chapter_2011,weigel_ensemble_chapter_2011}.
In the present paper, we will be concerned with {\em reliability}.
A formal definition (in the context of ensemble forecasts) will be given in Section~\ref{sec:serial_dependence}, but roughly speaking, an ensemble forecasting system is reliable if at any point $n$ in time, the ensemble members $X_1(n), \ldots, X_K(n)$ and the verification $Y(n)$ can be considered as having been drawn independently from an underlying (or latent) forecast distribution.
Reliability can be regarded as a statistical null hypothesis, and the aim of this paper is to develop tests for this null hypothesis.
In essence, this means to check whether the null hypothesis is plausible given actual data, that is, an archive of verifications and corresponding ensemble forecasts.
A popular tool to assess the reliability of ensemble forecasting systems are rank histograms~\citep[see e.g.][]{anderson96,hamill97,talagrand97,hamill01}.
It is assumed that the verifications are real numbers; it is therefore possible to determine, for any time instant $n$, the rank $R(n)$ of the verification $Y(n)$ among the ensemble members $X_1(n), \ldots, X_K(n)$.
The rank $R_n$ can assume the values $1, \ldots, K+1$, and if the ensemble forecasting system under concern is reliable, the distribution of $R_n$ is uniform over these values.
This implies that a reliable ensemble forecasting system should produce a ``more or less'' uniform rank histogram. 
In reality a rank histogram will never be precisely flat, and there are broadly speaking two possible reasons for this. 
Firstly, deviations from the uniform distribution might be due to the ensemble forecasting system failing to be reliable.
There are certain deficiencies of ensemble forecasting systems that appear to be somewhat typical and which produce characteristic patterns in the rank histogram.
A U-shaped distribution for instance indicates underdispersiveness, with a peaked distribution suggesting the opposite; sloped rank histograms show under-- or overforecasting (depending on the sign of the slope).
Secondly, even a perfectly reliable ensemble forecasting system will not produce a perfectly uniform rank histogram due to random variations.
Thus a test for reliability essentially amounts to a test for the hypothesis that the ranks have a discrete uniform distribution.
A common test for evaluating whether a histogram is consistent with a specific discrete distribution is Pearson's goodness--of--fit (GOF)~test.
(Taking the ordering of the possible ranks into account, which the GOF~test does not, more powerful tests can be obtained, for instance from the Cram\'{e}r--von~Mises family of statistics, see~\citet{elmore05}. In the present paper, we will focus on variants of the GOF~test though.)
A serious problem with applying the GOF~test directly to rank histograms for ensemble forecasting systems though is that the ranks are generally not independent. 
This will be demonstrated in Section~\ref{sec:numerical_examples} with a simple toy example.
Independence however is an important assumption in the GOF~test that can not easily be dispensed with.
The general fact that verification--forecast pairs can certainly not assumed to be independent is a difficulty that affects statistical forecast evaluation in general, as has been emphasised only relatively recently~\citep[see for instance][]{wilks_sampling_distributions_brier_score_2010,siegert_forecast_correlation_skill_2017,broecker_framework_dependence_2018}. 
A remedy suggested by~\citet{wilks_sampling_distributions_brier_score_2010} is to use explicit
(parametric) assumptions regarding the dependence structure and distribution of the
forecasts, but the considered situation is very specific.
In the present paper, we will use an approach based on results similar to~\citet{broecker_kuna_forecasting_systems_2018,broecker_framework_dependence_2018}.
The basic idea is that assuming the forecasting system is reliable, the ensemble $\bvc{X}(n) = (X_1(n), \ldots, X_K(n))$ provides the statistical properties of $Y(n)$, given the information available at the time the forecast $\bvc{X}(n)$ was issued, namely at time $n - L$, where $L$ is the lead time.
This fact can be used to obtain (to some extent) the statistical properties of the ranks, including their correlation structure.
In fact, in certain cases (corresponding effectively to lead time $L = 1$) the ranks turn out to be independent after all, meaning that in this situation the classical GOF~test can be used. 
In general though, the more complicated correlation structure of the ranks needs to be taken into account. 
We will show that this is possible, however.
By modifying GOF--like tests in an appropriate manner, we obtain tests for the reliability of ensemble forecasts.
These tests are valid under fairly general extraneous assumptions (by which we mean assumptions that would not be shared by {\em all} reliable ensemble forecasting systems).
\section{The goodness--of--fit test revisited}
\label{sec:gof_revisited}
In this section, we will revisit the basic steps in deriving the distribution of the goodness-of-fit test statistic.
In particular, we will clarify where the assumption of independence of the ranks comes in.
We start with fixing some general notation.
We let $\{Y(n), n = 1, \ldots, N\}$ be a series of real--valued verifications, with the index $n$ representing the time.
Further, $\{\bvc{X}(n), n = 1, \ldots, N\}$ is a series of corresponding ensemble forecasts, where for each time instant $n$ the ensemble is given by a vector of $K-1$ ensemble members, that is $\bvc{X}(n) = (X_1(n), \ldots, X_K(n))$, where each ensemble member is again real valued.\footnote{Using $K-1$ rather than $K$ ensemble members will simplify subsequent notation.} 
For a given $y \in \R$ and $\bvc{x} \in \R^{K-1}$, we consider the function $r(y, \bvc{x})$ that is equal to $k$ if the rank of $y$ among the $K$--dimensional vector $(y, \bvc{x})$ is equal to $k$.
In other words, $r(y, \bvc{x}) = k$ if precisely $k-1$ components of $\bvc{x}$ are smaller than or equal to $y$.
The function $r$ can assume the values $1, \ldots, K$.
For $n = 1, \ldots, N$, we define $R(n) := r(Y(n), \bvc{X}(n))$, that is $R(n)$ is the rank of the verification $Y(n)$ with respect to the ensemble $\bvc{X}(n)$.
We assume that the ensemble forecasting system is {\em reliable} with respect to the verifications. 
As said in the introduction, this means broadly speaking that for each time $n$, the verification $Y_n$ as well as each individual ensemble member $X_k(n), k = 1, \ldots, K-1$ can be considered independently drawn from some underlying forecast distribution.
This implies (again, a proof will follow in the next section) that for each $n$ the rank $R_n$ is uniformly distributed over its possible values $\{1, \ldots, K\}$.
As has already been mentioned though, there is no apriori reason why the ranks $R(n), n = 1, 2, \ldots$ should be independent from one another.
To define the GOF~test~statistic, consider the counts
\beqn{equ:2.10}
N_k := (\text{Number of $n$ for which $R(n) = k$})
 = \sum_{n = 1}^{N} \cf_{\{R(n) = k\}},
\eeq
where the {\em indicator function} $\cf_{A}$ of some event $A$ is one if the event happens and zero otherwise, and $k = 1, \ldots, K$.
Clearly, the count $N_k$ is the height of the $k$'th histogram bar.
Further we set 
\beqn{equ:2.20}
c_k := \frac{N_k - N/K}{\sqrt{N/K}}.
\eeq
Note that the expected value of $c_k$ is zero, since $N/K$ is the expected number of counts for each value of the rank, or alternatively the expected height of the $k$'th histogram bar.
The GOF~test~statistic is given by 
\beq{equ:2.30}
t = \sum_{k = 1}^K c_k^2 = \|\bvc{c}\|^2, 
\eeq
where $\bvc{c}  = (c_1, \ldots, c_K)$ and $\|.\|$ denotes the standard Euclidean norm.
The test statistic $t$ has, asymptotically for large $N$, a $\chi^2$~distribution with $K-1$ degrees of freedom, if the ranks are indeed independent. 
This can be seen as follows.
The key property of the variables $c_1, \ldots, c_K$ is that they jointly satisfy a central limit theorem; for this to happen, it is sufficient that the ranks $R(n), n = 1, \ldots, N$ are independent. 
It is worth noting already at this point though that independence is not necessary, as will be discussed in the next section.
In any event, we assume that the $c_1, \ldots, c_K$ have a joint normal distribution, with mean zero as was already noted.
We now have to calculate the covariance matrix, but before doing this, we note the following fact: let $\bvc{v} \in \R^{K}$ be the vector with components $v_k = 1/\sqrt{K}$ for all $k = 1, \ldots, K$.
Then $\|\bvc{v}\| = 1$ and also
\beqn{equ:2.35}
\bvc{v}^T \bvc{c} = \sum_{k = 1}^K c_k v_k = \frac{1}{\sqrt{K}} \sum_{k = 1}^K c_k = 0.
\eeq
If we now write $\Gamma_{i, j} := \E (c_i c_j)$ for the covariance matrix of $\bvc{c}$, then 
\beqn{equ:2.40}
(\Gamma \bvc{v})_i
 = \sum_{j = 1}^K \E (c_i c_j) v_j
 = \E (c_i \sum_{j = 1}^K c_j v_j)
 = 0. 
\eeq
This means that the nullspace (or kernel) of $\Gamma$ is spanned by the constant vecor $\bvc{v}$; we stress that this is true irrespective of whether the ranks are independent or not.
To find the precise shape of the covariance matrix $\Gamma$ though, we have to use independence. 
A simple calculation will then reveal that
\beq{equ:2.50}
\Gamma = \cf - \bvc{v} \cdot \bvc{v}^T.
\eeq
This matrix is symmetric and has a nullspace spanned by $\bvc{v}$ (as was already seen), while any other vector $w$ with the property that $\bvc{v}^T \bvc{w} = 0$ is an eigenvector of $\Gamma$ with eigenvalue one.
The condition that $\bvc{w}$ is perpendicular to $\bvc{v}$ just means that $\sum_{k = 1}^K w_k = 0$; vectors with this property are called {\em contrasts}.
Let now $\bvc{w}^{(1)}, \ldots, \bvc{w}^{(K-1)}$ be a set of orthogonal contrasts (such a set can contain at most $K-1$ elements).
Then the random variables $\bvc{d} = (d_1, \ldots, d_{K-1})$ defined through
\beq{equ:2.60}
d_j  = \sum_{k = 1}^K c_k w_k^{(j)}
\eeq
have again a normal distribution with mean zero, but now with unit covariance matrix, since $\E(d_j d_k) = (\bvc{w}^{(j)})^T \Gamma \bvc{w}^{(k)} = \delta_{jk}$.
It follows that $d_1, \ldots, d_{K-1}$ are independent standard normal. 
Therefore, $\sum d_k^2$, where the index $k$ runs over a subset of $\{1, \ldots, K-1\}$, has a $\chi^2$~distribution, with degrees of freedom given by the size of that subset.
In particular, $\|\bvc{d}\|^2$ has a $\chi^2$~distribution with $K-1$ degrees of freedom.
But since $\|\bvc{d}\|^2 = \|\bvc{c}\|^2 = t$, the same is true for $t$.
As an aside, we note that a user has the option to assess the rank histogram by using only a subset of the random variables $d_1, \ldots, \ldots, d_{K-1}$, or in other words, by projecting the scaled counts $c_1, \ldots, \ldots, c_{K}$ onto a reduced set of contrasts.
This has been suggested previously by~\citet{jolliffe_chi2_decomposition_2008}.
The user has complete freedom in choosing the desired contrasts, as long as they are orthogonal and normalised.
To obtain such a set, it is suggested to start with a set of vectors $\bvc{u}^{(1)}, \ldots, \bvc{u}^{(\kappa)}$ that have roughly the desired shape (for instance linear, U--shaped, sinusoidal, etc) and then apply a Gram--Schmidt procedure (or $QR$--decomposition) to the vectors $\bvc{v}, \bvc{u}^{(1)}, \ldots, \bvc{u}^{(\kappa)}$.
\section{Tests valid under serial dependence}
\label{sec:serial_dependence}
In the previous section, we discussed why the classical GOF~test~statistic has a $\chi^2$~distribution with $K-1$ degrees of freedom.
If we look back at this discussion, we find that the independence of the ranks was used in two places: in justifying a Central Limit Theorem for the $c_1, \ldots, c_K$, and when calculating the precise form of the covariance matrix $\Gamma$.
With the condition of independence dropped, $\Gamma$ will not have any longer the form shown in Equation~\eqref{equ:2.50}, and this is the main reason why applying the standard GOF~test to rank histograms is not warranted in general.
We will discuss later in this section that a Central Limit Theorem might still hold even though the ranks are not independent.
Further, even though $\Gamma$ is no longer known, the relevant correlations can be estimated from the data, and an estimator will be provided below.
For now, we assume that these random variables have a normal distribution with mean zero and some covariance matrix $\Gamma$.
It remains true though that the nullspace of $\Gamma$ is spanned by the vector $\bvc{v}$ as the derivation of this fact in the previous section did not depend on independence of the ranks.
This implies that we still get a faithful representation of the scaled counts $c_1, \ldots, c_{K}$ by projecting then onto a set of orthonormal contrasts as in Equation~\eqref{equ:2.60}, that is by using the random variables $d_1, \ldots, d_{K-1}$ defined through Equation~\eqref{equ:2.60}.
We want to develop a test based on a subset $\bvc{d} = (d_1, \ldots, d_{\kappa})$ of these random variables, and we denote the covariance matrix of these random variables by $\Upsilon_{i, j} = \E(d_i d_j) = (\bvc{w}^{(i)})^T \Gamma \bvc{w}^{(j)}$, where $i, j \leq \kappa \leq K-1$.
We keep $\kappa$ fixed throughout the remainder of this section.
As the the condition of independence of the ranks has been dropped, $\Upsilon$ will not be the unit matrix any longer.
(We note again that $\Upsilon$ will later have to be estimated from the data.)
We consider the statistic $t_{\kappa} = \bvc{d}^T \Upsilon^{-1} \bvc{d}$.
This statistic is indeed a generalisation of the statistic $t$ from the previous section, and the two agree if the ranks are independent and $\kappa = K-1$.
Our claim is that $t_{\kappa}$ has a $\chi^2$--distribution with $\kappa$ degrees of freedom as in the independent case.
To see this, let $U$ be a symmetric matrix so that $U \Upsilon U =  \cf $ (i.e.\ $S$ is a square root of $\Upsilon^{-1}$).
Then $\bvc{e} = U \bvc{d}$ is a vector of normal random variables with zero mean and covariance matrix $U\Upsilon U = \cf$, hence the components of $\bvc{e}$ are independent and standard normal.
As a consequence, $\tilde{t} = \|\bvc{e}\|^2$ has a $\chi^2$--distribution with $\kappa$ degrees of freedom.
However, 
\beqn{equ:3.20}
\tilde{t} = \|\bvc{e}\|^2 = \bvc{e}^T \bvc{e} = \bvc{d}^T U \cdot U \bvc{d} = \bvc{d}^T \Upsilon^{-1} \bvc{d} = t_{\kappa},
\eeq
proving our claim.
For the remainder of this section, we will fill in the missing parts of our argument.
We will show that although the ranks are not independent, they nevertheless satisfy a very strong decay of correlation property which is a direct consequence of the reliability assumption and forms the core of our analysis. 
We then provide an estimator of the covariance matrix $\Upsilon$.
The feasibility of this estimator is due to the strong decorrelation property of the ranks, and the assumption that the ranks form an ergodic sequence; this is the only extraneous assumption we need to add. 
These properties are also sufficient to justify the validity of the Central Limit Theorem (more details will be provided in Appendices~\ref{apx:covariance} and~\ref{apx:clt}).
The reliability assumption is interpreted to mean the following.
For every time instant $n = 1, \ldots, N$ there exists an underlying or latent forecast distribution $\mu_n$ over the real numbers.
This distribution is itself random and given by the distribution of $Y_n$ conditional on the information available at forecast time.
More formally, let $\cF_n$ be the information available to the forecaster at time $n$, and say that forecasts are issued with a lead time $L$, then reliability means that
\beqn{equ:3.40}
\mu_n(A) = \P(Y(n) \in A | \cF_{n-L})
\eeq
for all $n = 1, \ldots, N$ and any set $A$ on the real line.\footnote{%
Strictly speaking for any measurable set $A$ on the real line.
}
The joint set of verification and ensemble members $(Y_n, X_1(n), \ldots, X_{K-1}(n))$ are independently drawn from this distribution, that is, for any $n$ and any sets $A_0, \ldots, A_{K-1}$ on the real line, it holds that
\beqn{equ:3.50}
\begin{split}
& \P(Y(n) \in A_0,  X_1(n) \in A_1, \ldots, X_{K-1}(n) \in A_{K-1}| \cF_{n-L}) \\
& = \mu(n, A_0) \cdot \ldots \cdot \mu(n, A_{K-1}).
\end{split}
\eeq
The uniform distribution of the ranks, {\em conditional} on the forecast information, is now an elementary consequence: for all $n = 1, \ldots, N$ and $k = 1, \ldots, K$ we have
\beq{equ:3.60}
\P(R(n) = k| \cF_{n-L}) 
= \frac{1}{K}.
\eeq
We will graft another element to the reliability assumption which is usually not made explicit but is evidently satisfied in most applications, namely that
for any $n$, the forecast information $\cF_n$ contains all verifications and ensembles up to that point; in other words, at any time $n$ the forecaster knows $\{Y(m), m = 1, \ldots, n\}$ and also $\{\bvc{X}(m), m = 1, \ldots, n\}$.
This, in combination with Equation~\eqref{equ:3.60}, yields the following key identity:
\beq{equ:3.70}
\P(R(n) = k| R(1), \ldots, R(n-L)) 
= \frac{1}{K}
\eeq
for all $n = 1, \ldots, N$ and $k = 1, \ldots, K$.
Another way of saying this is that for any $n$, the rank $R_n$ is uniformly distributed and independent from the ranks $R(1), \ldots, R(n-L)$, that is, from the ranks known at forecast time. 
In particular, we obtain that in the case of unit lead time (i.e.~$L = 1$), the ranks $\{R(n), n = 1, 2, \ldots\}$ are indeed fully independent; this implies that in this special (but important) situtation, the classical GOF~test for the rank histogramm is valid.
Let now $\{\bvc{w}^{(1)}, \ldots, \bvc{w}^{(\kappa)} \}$ be a set of orthonormal contrasts, and define 
\beq{equ:3.80}
Z_k(n) = \sqrt{K} \sum_{j = 1}^{K} w^{(k)}_j\cf_{\{R(n) = j\}}
\eeq
for $n = 1, \ldots, N$ and $k = 1, \ldots, \kappa$; note that $d_k = \frac{1}{\sqrt{N}} \sum_{n = 1}^N Z_k(n)$.
We regard $\bvc{Z}(n) = (Z_1(n), \ldots, Z_{\kappa}(n))$ with $n = 1, \ldots, N$ as a sequence of random vectors.
The property~\eqref{equ:3.70} implies that this sequence has finite correlation length $L-1$.
To see this, note that for any $n$, the random vector $\bvc{Z}(n)$ depends on $R(n)$ only, and further that $\E(\bvc{Z}(n)) = 0$.
Hence, $\bvc{Z}(n+l)$ is independent of $\bvc{Z}(n)$ if $l \geq L$, and we have 
\beq{equ:3.90}
\E(\bvc{Z}(n+l) \cdot \bvc{Z}(n)^T) 
 = \E(\bvc{Z}(n+l)) \cdot \E(\bvc{Z}(n)^T)
 = 0.
\eeq
The first equality follows because $\bvc{Z}(n)$ is a function of $R(n)$ and hence also a function of $R(1), \ldots, R(n+l-L)$ (since $l \geq L$), the second inequality follows from property~\eqref{equ:3.70}, and the third from $\E(\bvc{Z}(n)) = 0$.
It turns out that in order to establish a joint Central Limit Theoreman for $\bvc{d} = (d_1, \ldots, d_{\kappa})$, an additional assumption is needed, namely that the ranks $\{R(n), n = 1, 2, \ldots \}$ form a stationary and ergodic sequence. 
With this assumption and property~\eqref{equ:3.90} in place, it follows from existing results that $\bvc{d}$ will be asymptotically normal with mean zero and some covariance matrix $\Upsilon$; we will not provide a proof here, but some more details and references can be found in Appendix~\ref{apx:clt}.
An estimator for $\Upsilon$, the asymptotic covariance matrix of $\bvc{d}$, is needed as well.
We will use the estimator
\beq{equ:3.100}
\Upsilon_N = \cf + \frac{1}{N} \sum_{n = 1}^{N} \sum_{l = 1}^{L-1} \bvc{Z}(n) \bvc{Z}(n+l)^T + \bvc{Z}(n+l) \bvc{Z}(n)^T. 
\eeq
This estimator can be shown to converge to $\Upsilon$, and a demonstration can be found in Appendix~\ref{apx:covariance}.
We stress that the validity of this estimator rests not only on the ergodicity assumption but also on the finite correlation property~\eqref{equ:3.90}.
For the case $L = 1$, this estimator reduces to $\Upsilon_N = \cf$ as it should.
\section{Numerical examples}
\label{sec:numerical_examples}
\begin{figure}[p]
\begin{center}
\includegraphics[width = \columnwidth]{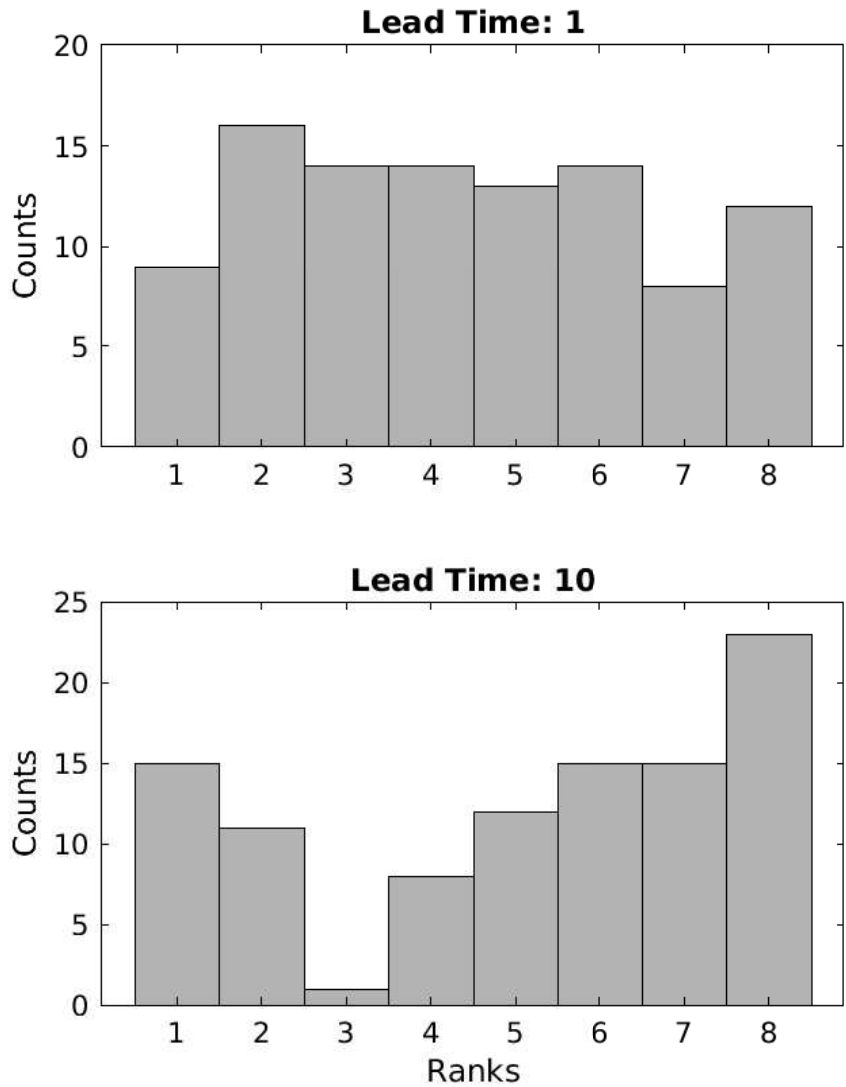}
\caption{\label{fig:simple_ensemble_typical_hist} Typical histograms for ensemble forecasts for the AR~process.
The ensemble had 7~members, and the data set comprised 100~time instances.
The lead time was 1~time unit for the top panel and 10~time units for the bottom panel.
Althoug both forecast systems are by construction reliable, the histogram for the larger lead time is considerably ``rougher'', that is there are stronger variations in the counts.
This is due to the positive temporal correlations between the ranks for the forecasting system at larger lead times.}
\end{center}
\end{figure}

\begin{figure}[p]
\includegraphics[width = \columnwidth]{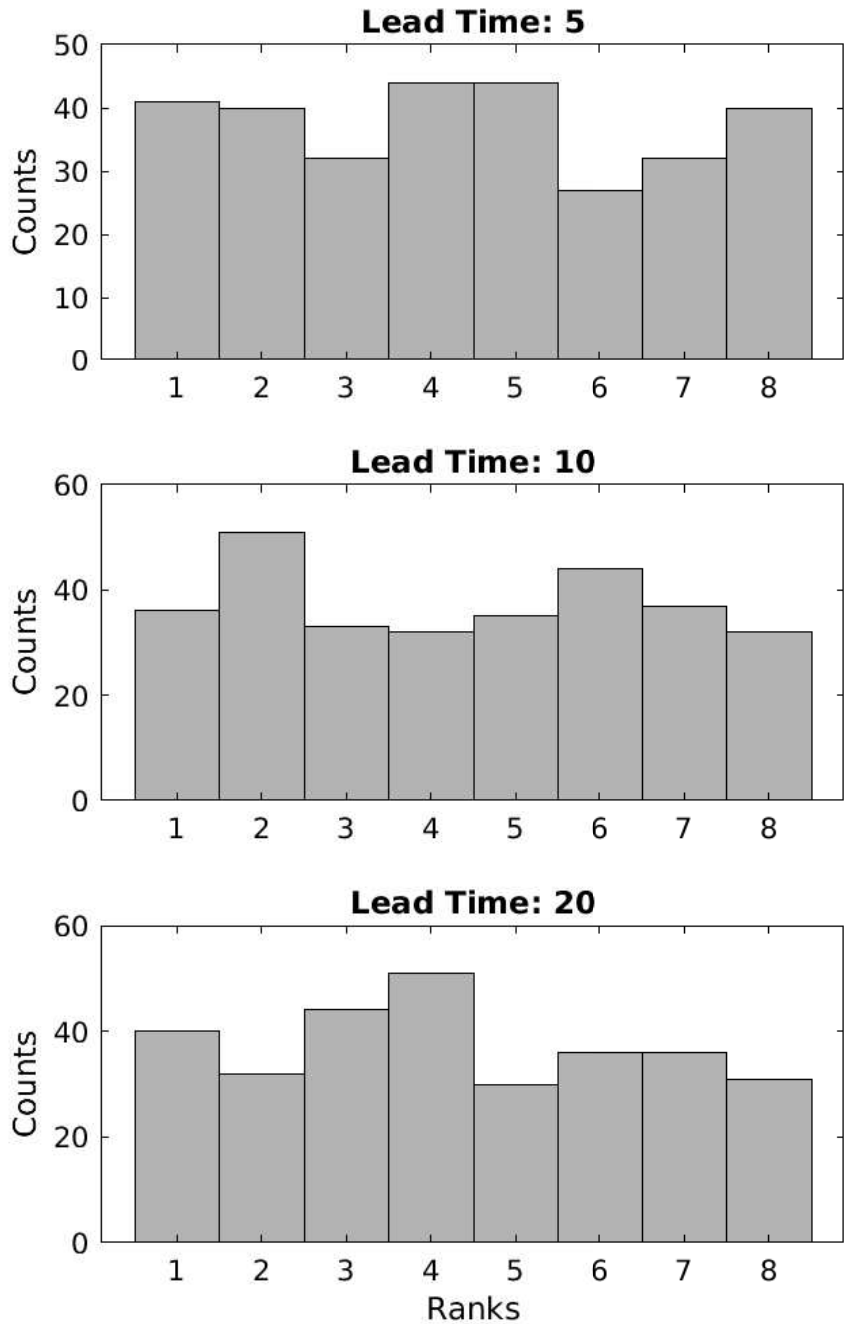} 
\caption{\label{fig:bve_histograms}Rank histograms for lead times of $L = 5$,~$10$ and~$20$ in the Navier--Stokes experiment (top, middle, and bottom panel, respectively).
Each data set comprised $300$~verification--forecast pairs.
There is no obvious deviation from reliability, although the histogram for lead time $L=20$ might be slightly slanted to the right by visual inspection. 
The test detects no significant deviation from reliability though.}
\end{figure}

We start this section with a short list summarising the steps needed to perform the test for flatness of a rank histogram.
We let $\{(Y(n), \bvc{X}(n)), n = 1, \ldots, N\}$ be a sequence of real--valued verifications and corresponding ensembles with $K-1$ members.
Let further $\{\bvc{w}^{(1)}, \ldots, \bvc{w}^{(\kappa)}\}$ be a set of orthonormal contrasts, describing possible deviations of a rank histogram from flatness (with $\kappa \leq K-1$).
\begin{enumerate}
\item Compute the ranks $\{R(n), n = 1, \ldots, N\}$.
\item Using the ranks and the contrasts, compute $Z_k(n)$ from Equation~\eqref{equ:3.80} for $n = 1, \ldots, N$ and $k = 1, \ldots, \kappa$.
\item Compute the estimator $\Upsilon_N$ for the covariance $\Upsilon$ from Equation~\eqref{equ:3.100}.
\item Compute $d_k = \frac{1}{\sqrt{N}} \sum_{n = 1}^N Z_k(n)$ for $k = 1, \ldots, \kappa$ and let $\bvc{d} = (d_1, \ldots, d_{\kappa})$.
\item Now $\bvc{d}^T\Upsilon_N^{-1}\bvc{d}$ should have a $\chi^2$~distribution with $\kappa$ degrees of freedom, and this can be used to compute the $p$--value.
\end{enumerate}
For the remainder of this section, we will discuss two numerical examples.
The first example considers a simple autoregressive process; this has been chosen merely to illustrate the methodology.
The second example uses data from an assimilation experiment using the two~dimensional Navier--Stokes equation.
%
%%%%%%%%%%%%%%%%%%%%%%%%%%%%%%%%%%%%
%
\paragraph{\bfseries Example~1: Autoregressive process}
In the first example, the verification $\{Y_n, n = 1, 2, \ldots \}$ forms an autoregressive (AR)~process of the form
\beq{equ:4.10}
Y(n+1) = \alpha Y(n) + \zeta(n+1),
\eeq
where $\{\zeta(n), n \in \Z \}$ is a sequence of independent standard normal random variables and $\alpha = 0.95$.
The information $\cF_n$ available to the forecaster at time $n$ is $\{Y(k), k \leq n \}$, that is the entire history of observations up to and including $Y(n)$.
Reliable ensemble forecasts can be generated by replacing $\zeta(n)$ in Equation~\eqref{equ:4.10} with independent realisations of the noise process.
More specifically, let $\{\pmb{\xi}(n), n = 1, 2, \ldots \}$ be a sequence of independent random vectors $\pmb{\xi}(n) = (\xi_1(n), \ldots, \xi_{K-1}(n))$, where the components $\xi_k(n)$ are again independent and standard normal.
Then an ensemble forecast for lead time $L$ and verifying at time $n+L$ is given by 
\beqn{equ:4.20}
Z^{k}(n+L) = \alpha^L Y(n) + \sigma_L \xi_k(n),
\qquad k = 1, \ldots, K-1;
\eeq
here, $\sigma_L^2 = \sum_{l = 0}^{L-1} \alpha^{2l}$.
In this model, it is easy to see directly that two ranks are independent if they are $L$ or more steps apart, but that they are dependent otherwise.
To check this, we write $Y(n+L)$ as 
\beq{equ:4.30}
Y(n+L) = \alpha^L Y(n) + \sum_{l = 0}^{L-1} \alpha^{l} \zeta(n+L-l).
\eeq
Therefore, 
\beq{equ:4.40}
\begin{split}
R(n+L) & = r(Y(n+L), \bvc{Z}(n+L))\\
 & = r(\sum_{l = 0}^{L-1} \alpha^{l}\zeta(n+L-l), \sigma_L \pmb{\xi}(n)).
\end{split}
\eeq
(We recall that $r(y, \bvc{x})$ is the rank of $y$ among the components of $\bvc{x}$.)
Equation~\eqref{equ:4.40} demonstrates that the temporal dependence of the ranks is due to the temporal dependence of $\iota_L(n) := \sum_{l = 0}^{L-1} \alpha^{l}\zeta(n+L-l)$.
In view of Equation~\eqref{equ:4.30}, the random variable $\iota_L(n)$ describes the subsequent evolution of the observations after the forecast $\bvc{Z}(n)$ has been issued.
We might call $\iota_L(n)$ the {\em innovation}; it is precisely the part of $Y(n+L)$ not captured by the forecast.
If two observations $Y(n)$ and $Y(m)$ are less than $L$ time steps apart (i.e.\ $|m - n| < L$), then their corresponding innovations will be dependent, due to overlap of their evolutions after the respective forecasts have been issued.
This is also evident from the expression of the innovation.
If $|m - n| \geq L$ though, their innovations will be independent.
Due to Equation~\eqref{equ:4.40}, the ranks will exhibit the same phenomenon.
Figure~\ref{fig:simple_ensemble_typical_hist} shows typical histograms for ensemble forecasts in the context of the AR~process.
The ensemble forecasting system uses 7~members, and the data set comprised 100~time instances.
The lead time was 1~time unit for the top panel and 10~time units for the bottom panel of Figure~\ref{fig:simple_ensemble_typical_hist}.
It is evident that the histogram for the larger lead time shows considerably stronger variations in the counts.
This is due to the positive temporal correlations between the ranks at larger lead times.
The $p$--values for the top and bottom panels are~0.7612 and~0.7199, respectively, using the test proposed in Section~\ref{sec:serial_dependence} for the second histogram.
Using a classical GOF~test would give a $p$--value of~0.0019 for the second histogram, thus concluding wrongly that this forecast is not reliable.
In order to check whether the test presented in Section~\ref{sec:serial_dependence} takes the correlations correctly into account, we have created 1,000~Monte~Carlo resamples of the experiment described above, albeit with 400~time instances.
For every Monte~Carlo sample, we computed the statistic $t_{\kappa}$ for $\kappa = 2$, using a linear and a U--shaped contrast, as described in Section~\ref{sec:serial_dependence}, including the estimator of the covariance matrix.
\begin{figure}[p]
\begin{center}
\includegraphics[width = \columnwidth]{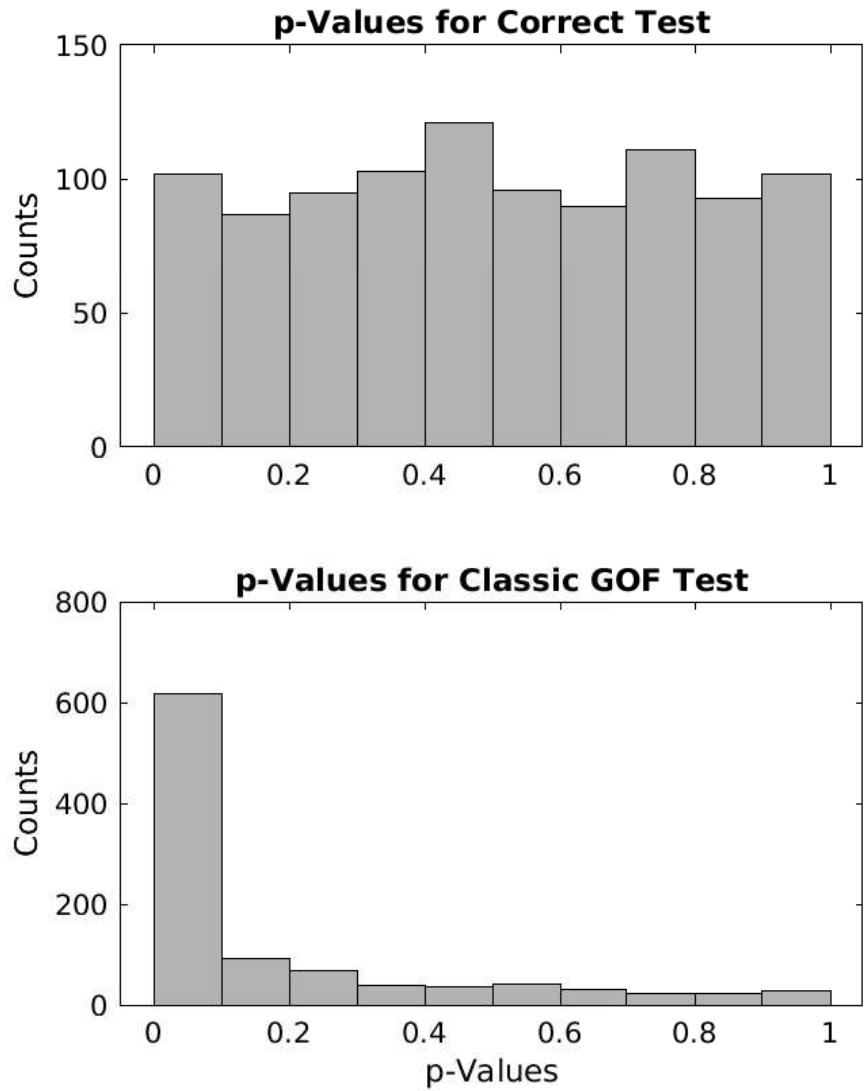}
\caption{\label{fig:simple_ensemble_mcsim} Histograms of the $p$--values of testing flatness of the rank histograms for the AR~process at lead time~10.
The ensemble had 7~members, and the data set comprised 400~time instances.
The test statistic employed two contrasts (linear and U--shaped).
The $p$--values were obtained from~1000 Monte Carlo repetitions of the same experiment.
The top panel shows the $p$--values from the new test proposed in Section~\ref{sec:serial_dependence} taking the rank correlations into account.
The bottom panel shows the $p$--values from a classical GOF~test.
It can be seen that the new test produces correct $p$--values, while ignoring the rank correlation results in too low $p$--values and thus too frequent rejection of the null hypothesis.}
\end{center}
\end{figure}
If the presented theory is correct, then $t_{\kappa}$ should follow a $\chi^2$ distribution with~$2$ degrees of freedom, or equivalently the $p$--value should have a uniform distribution.
This turns out to be the case; a histogram of the $p$--values obtained from our 1,000~Monte~Carlo resamples is shown in the top panel of Figure~\ref{fig:simple_ensemble_mcsim}.
Furthermore, a Kolmogorov--Smirnov test yields a $p$--value of~0.6876, confirming that these follow a uniform distribution.
For each Monte~Carlo resample we have also calculated the {\em classical} GOF statistic, that is, ignoring the correlations in the ranks and assuming that $\Upsilon$ is the identity matrix.
That the resamples of that statistic do {\em not} follow a $\chi^2$ distribution with~$2$ degrees of freedom is evident from the bottom panel of Figure~\ref{fig:simple_ensemble_mcsim}, which shows a histogram of the $p$--values.
These are evidently concentrated at too low values, which implies that ignoring the correlations in the ranks and applying the classical GOF~test would result in too frequent rejection, that is, we would conclude too often that the rank histogram is not consistent with reliability.
% 
%%%%%%%%%%%%%%%%%
%
\paragraph{\bfseries Example~2: Data assimilation in 2D~Navier--Stokes}
The second example uses data from an assimilation experiment with the two~dimensional Navier--Stokes equation.
The equation was implemented in the vorticity--streamfunction formulation
\beq{equ:4.100}
\pdd_t \omega + J(\omega, \psi) + A \omega = f,
\eeq
on the two--dimensional unit torus $\mathbb{T} = ]0, 1[^2$ with periodic boundary conditions.
Here, $\omega$ is the vorticity and $\psi$ the stream function; further, $A = -\nu \Delta$ (the Laplacian with viscosity $\nu$), and the stream function is obtained from the vorticity through solving the Poisson equation $\Delta \psi = \omega$.
The function $f$ represents a forcing.
Equation~\eqref{equ:4.100} (along with the Poisson equation) was solved with a pseudospectral code on a square spatial lattice with resolution $N = 21$ in both dimensions.
In other words, the equation was truncated at wavenumber~10, where we define the {\em wavenumber} of a wave vector $(k, l)$ as $|(k, l)| := \max \{|k|,|l|\}$.
The viscosity was set to $\nu = 2\cdot 10^{-3}$.
The forcing was time independent and composed of randomly selected amplitudes and truncated at wavenumber~3, with a magnitude of $\|f\| = 1.34$.
(Here and in the following, we use the norm $\|f\| = \left(\int_{\mathbb{T}} |f|^2(x) \dd x \right)^{1/2}$ for a---possibly complex---function on the torus.) 
The data was assimilated into an identical copy of the two~dimensional Navier--Stokes equation.
As observations, the Fourier modes with wavenumbers $|(k, l)| \leq 1$ were used (which corresponds to observing nine modes, or equivalently, to taking smoothed spatial observations on a grid with $3 \times 3$ gridpoints).
The observations were taken at temporal intervals of $\Delta t = 0.5$ time units and corrupted with normally distributed noise of about $5\%$.
The observations were then assimilated simply by replacing the relevant Fourier modes of the assimilated solutions with the observed Fourier modes~\citep[see][for theoretical analyses of this assimilation method]{hayden_discrete_DA_navierstokes_2011,sansalonso_filterasymptotics_2014,oljaca_almost_sure_error_2017}.
Ensembles were generated by randomly perturbing the analyses fields.
The average size of the perturbing fields $\delta \omega$ was set to $\|\delta \omega\| = 0.943$ for all lead times; this value was found by optimising the mean square forecast performance for lead time of~$5$ units in an offline experiment. 
We analysed ensembles for lead times of $L = 5$,~$10$ and~$20$ time units, each data set comprising $300$~verification--forecast pairs.
The histograms for these three data sets are shown in Figure~\ref{fig:bve_histograms}.
No attempt was made to statistically recalibrate these ensembles.
It is seen that the reliability of this relatively simple ensemble forecasting system is not bad by visual inspection.
We applied the described test for flatness of the rank histogram, first for two contrasts (i.e.\ $\kappa = 2$).
The $p$--values for lead times $L = 5$,~$10$ and~$20$ are~0.7872,~0.7495, and~0.5209, respectively.
Testing the full set of contrasts gives $p$--values of~0.5507,~0.5572, and~0.5854; all these figures do not provide a strong case for deviation from reliability. 
With regards to the rank histogram corresponding to lead time $L = 20$ in particular though, the histogram appears to have a slight slant to the right (indicating underforecasting), but this effect might be masked by the expected variation of the histogram.
However, we find that $\text{trace}(\Upsilon_N) = 8.63$, while this value would be~$7$ for independent ranks, and we can conclude that the variance of the histogram is not in fact much larger than for the independent case.
We repeated the test for lead time $L = 20$ with a single, linear contrast and find a $p$--value of~0.3254, which might indicate a slight deviation from reliability.
Note that we have cheated a little bit, as the choice of the contrast was made based on the data. 
As a final note, under the assumption of uncorrelated ranks the $p$--value for this case would have been~0.2676, so not in fact very different.
For the variance, we have the estimate $1.2708$ which is fairly close to~$1$, again indicating that dropping the assumption of independence does not make much of a difference in this case.
We stress however that there is no reason why this should be the case in any generality since the actual rank correlations are {\em not} universal and depend on the specific problem at hand.
% 
%%%%%%%%%%%%%%%%%%%%%%%%%%%%% 
% 
\section{Conclusions and outlook}
\label{sec:conclusions}
A popular and practical tool to assess the reliability of ensemble forecasting systems (for scalar verifications) is the rank histogram.
For a reliable ensemble forecasting system, this histogram is expected to be more or less flat, since the ranks are uniformly distributed among their possible outcomes.
For a more quantitative analysis though, it would be desirable to have a test for the flatness of rank histograms, as certain random fluctuations will always be present even if the forecasting system is reliable.
We have argued that classical approaches such as for example Pearson's goodness--of--fit test are not appropriate since these tests rest on the assumption that the ranks form a sequence of independent random variables.
By revising the derivation of Pearson's goodness--of--fit test, we identified two places where the assumption of independence is relevant: firstly it ensures that the rescaled histogram counts satisfy a joint Central Limit Theorem, and secondly it entails a very specific correlation structure for these counts. 
Although the ranks of a reliable ensemble forecasting system are not independent in general, we have demonstrated both analytically and numerically that an appropriate modification of the goodness--of--fit test will still work.
Central to our analysis is the fact that for a reliable ensemble forecasting system, the ranks still satisfy a strong decay of correlation property---the correlation time of the ranks is even finite and given by the lead time less one. 
(This result can be generalised to different types of forecasting systems and might be of independent interest, see~\citet{broecker_kuna_forecasting_systems_2018,broecker_framework_dependence_2018}.)
Furthermore, it was shown how to perform a ``reduced'' goodness--of--fit test using a restricted set of contrasts, as suggested in~\citet{jolliffe_chi2_decomposition_2008}, but modified so as to account for rank correlations.
Apart from the technical condition that the ranks form an ergodic sequence, the approach does not require any extraneous or distributional assumptions.
The formalism was also applied to numerical examples. 
First, data from a simple autoregressive process was considered, with ensemble forecasts that were by construction reliable.
The experiments confirm that the formalism gives the correct results, while not taking the rank correlations into account (by using a classical goodness--of--fit test) yields too high rejection rates as the distribution of the classical goodness--of--fit test statistic is no longer a $\chi^2$--distribution.
A second example used data from a simple fluid dynamical data assimilation experiment. 
The results show that despite a relatively crude data assimilation system, the ensembles are fairly reliable. 
We also addressed the question whether the test looses power for longer lead times as potentially systematic deviations from a flat rank histogram are masked by strong variability of the histogram counts, which seems not the case in that situation.
\paragraph{\bfseries Outlook and future work}
An important fact emerging from our analysis is that for a reliable ensemble forecasting system, the ranks exhibit a finite correlation time which cannot exceed the lead time. 
This result can be generalised to different types of forecasting systems as has been done in~\citet{broecker_kuna_forecasting_systems_2018,broecker_framework_dependence_2018}.
Strong decay of correlations though typically implies powerful asymptotic limit results such as Laws of Large Numbers and Central Limit Theorems.
It seems plausible that these can be exploited to analyse other forecast evaluation techniques rigorously under serial dependence; examples are reliability diagrams~\citep{broecker06-4} or Receiver (or Relative) Operating Characteristic~\citep{egan75,broecker_chapter_2011}.
An extension of the results in the present paper to {\em stratified} rank histograms would also be desirable~\citep{siegert_stratification_2011}. 
Stratified rank histograms provide a more detailed picture of reliability, conditional on different forecasting situations.
This extension seems to be fairly immediate and will be dealt with in a forthcoming paper.
\appendix
\section{Covariance estimator}
\label{apx:covariance}
In this appendix, we discuss an estimator for $\Upsilon$, the covariance matrix of $\bvc{d} = (d_1, \ldots, d_{K-1}) = \frac{1}{\sqrt{N}} \sum_{n = 1}^N \bvc{Z}(n)$ in the limit $N \to \infty$, that is
\beqn{equ:5.10}
\Upsilon = \lim_{N \to \infty} 
\frac{1}{N} \E\left[ 
    \left(\sum_{n = 1}^N \bvc{Z}(n) \right) 
    \left(\sum_{n = 1}^N \bvc{Z}(n) \right)^T 
    \right].
\eeq
(Notation and definitions are as in Sec.~\ref{sec:serial_dependence}.)
We start with studying the (matrix valued) covariance function
\beqn{equ:5.30}
\gamma(l) := \E(\bvc{Z}(n) \bvc{Z}(n + l)^T),
\eeq
noting that there is no dependence on $n$ since $\{\bvc{Z}(n), n = 1, 2, \ldots \}$ is assumed ergodic and thus in particular stationary; note also that $\gamma(l)$ is defined for negative $l$, too, and in fact $\gamma(-l) = \gamma(l)^T$.
Furthermore, we have $\gamma(l) = 0$ if $l \geq L$ due to Equation~\eqref{equ:3.90}.
An elementary calculation then gives
\beqn{equ:5.40}
\frac{1}{N} \E\left[ 
    \left(\sum_{n = 1}^N \bvc{Z}(n) \right) \left(\sum_{n = 1}^N \bvc{Z}(n) \right)^T 
    \right]
 = \sum_{l = -N + 1}^{N-1} (1 - \frac{|l|}{N}) \gamma(l)
\eeq
and hence
\beq{equ:5.50}
\Upsilon 
 = \lim_{N \to \infty} 
    \sum_{l = -N + 1}^{N-1} (1 - \frac{|l|}{N}) \gamma(l) 
 = \sum_{l \in \Z} \gamma(l).
\eeq
Thanks to Equation~\eqref{equ:3.90}, the sum in Equation~\eqref{equ:5.50} contains only finitely many nonzero terms, namely for $|l| < L$.
These terms can be estimated by empirical averages (i.e.~averages over time), that is 
\beqn{equ:5.60}
\gamma_N(l) = \frac{1}{N} \sum_{n = 1}^{N} \bvc{Z}(n) \bvc{Z}(n+l)^T,
\eeq
which converges to $\gamma(l)$ due to the condition that the ranks are ergodic (we only need estimators for $0 < l < L$ since $\gamma(-l) = \gamma(l)^T$ is symmetric and $\gamma(0)$ is the unit matrix).
The estimator $\Upsilon_N$ for $\Upsilon$ is given by replacing $\gamma(l)$ in Equation~\eqref{equ:5.50} with the estimators $\gamma_N(l)$.
This gives
\beq{equ:5.12}
\begin{split}
\Upsilon_N 
 & = \cf +  \sum_{l = 1}^{L-1} \gamma_N(l) + \gamma_N(l)^T \\
 &  = \cf + \frac{1}{N} \sum_{n = 1}^{N} \sum_{l = 1}^{L-1} 
    \bvc{Z}(n) \bvc{Z}(n+l)^T + \bvc{Z}(n+l) \bvc{Z}(n)^T.
\end{split}
\eeq
%
%%%%%%%%%%%%%%%%%%%%%%%%%%%%
%
\section{The Central Limit Theorem}
\label{apx:clt}
In this appendix, we justify the a joint Central Limit Theorem for $d = (d_1, \ldots, d_{K-1})$, where $d_k = \frac{1}{\sqrt{N}} \sum_{n = 1}^N Z_k(n)$.
By a classical argument known as the Cram\'{e}r--Wold device in probability theory~\citep[see for instance][~pg.16]{vaart_asymptotic_statistics_1998} it is sufficient to establish a central limit theorem for $\delta := \frac{1}{\sqrt{N}} \sum_{n = 1}^N \Lambda(n)$ where $\Lambda(n) := \pmb{\lambda}^T \bvc{Z}(n)$ for any vector $\pmb{\lambda} \in \R^{K-1}$, thereby reducing the problem from a vector~valued to a single~valued Central Limit Theorem.
Our assumptions and the discussion in the previous appendix entail that $\{\bvc{Z}(n), n = 1, 2, \ldots\}$ are ergodic and have summable correlations.
The same is therefore true for $\{\Lambda(n), n = 1, 2, \ldots\}$, and we can apply Theorem~4.18 in~\citet{vaart_time_series_2010} to conclude that the distribution of $\delta$ is asymptotically normal.
In summary, we obtain the required joint Central Limit Theorem for $(d_1, \ldots, d_{K-1})$.
\bibliographystyle{plainnat}
\bibliography{/home/pt904209/TeX/Literatur}
\end{document}